\documentclass[10pt,a4paper]{article}

\usepackage{times}

\usepackage{epsfig,wrapfig}
\usepackage{epstopdf} 
\usepackage{amsmath}
\usepackage{color}
\usepackage{graphicx}
\usepackage{caption}
\usepackage{subcaption}
\usepackage{array}
\usepackage{tabularx}
\usepackage{multirow}
\usepackage{gensymb}
\usepackage{fancyhdr}

\usepackage[lmargin=2.5cm, rmargin=2.5cm,tmargin=3.50cm,bmargin=3.50cm]{geometry}
\usepackage{indentfirst}

\usepackage{titlesec}
\titleformat{\section}[hang]
  {\centering}{\thesection}{1ex}{\normalsize \textsc}
\titleformat{\subsection}[hang]
  {}{\thesubsection}{1ex}{\normalsize \textit}

\newcommand{\acknowledgement}{\section*{\centering{\textnormal{\normalsize{\textsc{Acknowledgement}}}}}}

\renewcommand{\thesection}{ \normalsize \textnormal{\Roman{section}.}}
\renewcommand{\thesubsection}{\normalsize \textnormal{\textsc{\textit{\Alph{subsection}.}}}}

\def\e{\begin{equation}}
\def\f{\end{equation}}
\def\_#1{{\bf #1}}

\def\.{\cdot}

\begin{document}

\title{\large \textbf{Multi-Channel Reflectors: Versatile Performance Experimentally Tested}}
%
\def\affil#1{\begin{itemize} \item[] #1 \end{itemize}}
\author{\normalsize \bfseries \underline{S. N. Tcvetkova}, V. S. Asadchy, A. D\'{\i}az-Rubio, D.-H. Kwon, and S. A. Tretyakov
\\ 
}
\date{}
\maketitle
\thispagestyle{fancy} 
\vspace{-6ex}
\affil{\begin{center}\normalsize Aalto University, Department of Electronics and Nanoengineering, P.O. 15500, FI-00076, Aalto, Finland\\
svetlana.tcvetkova@aalto.fi
 \end{center}}

\begin{abstract}
\noindent \normalsize
\textbf{\textit{Abstract} \ \ -- \ \
We investigate multi-channel reflectors, such as a three-channel power splitter and a five-channel isolating mirror. These metasurface reflectors are able to control reflections from and into several directions while possessing a flat surface. We design, fabricate, and experimentally study these new devices, confirming that the performance is nearly perfect.
}
\end{abstract}

\section{Introduction}

Recently, thin composite layers, called {\it metasurfaces}, have proved their efficient performance for control and transformation of electromagnetic waves, which motivated our work  on new multi-channel metasurface devices. 
Reflectors with multi-functional response are of paramount importance for many applications,  from microwave to optical frequencies.
However, there has been no established method for engineering a single reflector, which would manipulate   waves incident from different directions in desired ways. 
A ``multi-channel device'',  in an analogy with microwave devices, means that we can independently control waves incident from several directions. A splitter is one of the well-known examples of three-port microwave devices, where the channels are input and output waveguides. Here, we discuss this paradigm for plane-wave ``ports''.  
Depending on the metasurface periodicity, plane waves may propagate along several ``open'' channels  \cite{multi_channel}, as follows from the Floquet theory. 
\begin{figure}[h!]
    \centering
    \begin{subfigure}[h]{0.2\textwidth}
        \includegraphics[width=\textwidth]{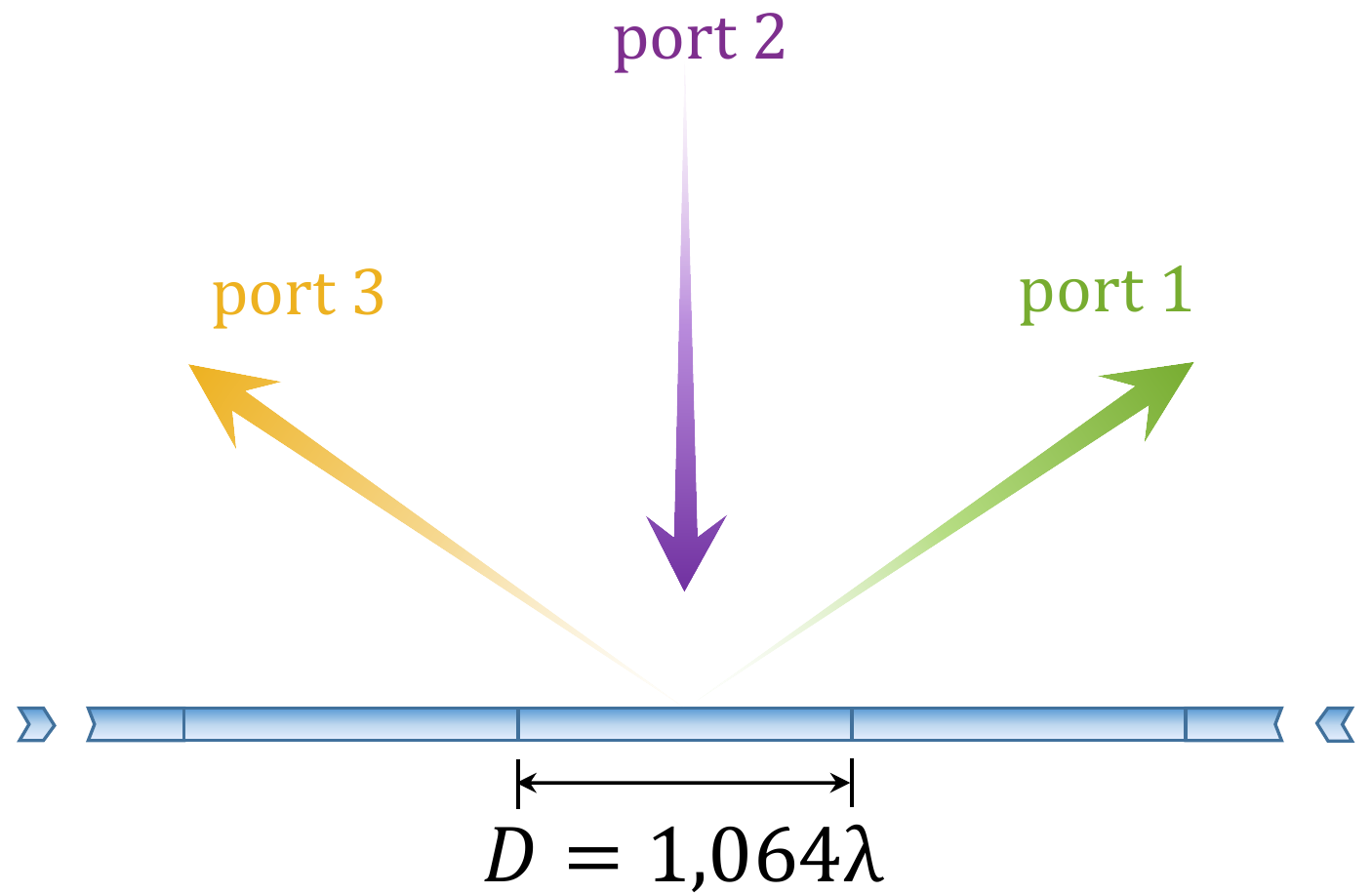}
          \caption{}
        \label{fig:1a}
            \end{subfigure}
                \hspace{1cm}
    \begin{subfigure}[h!]{0.2\textwidth}
        \includegraphics[width=\textwidth]{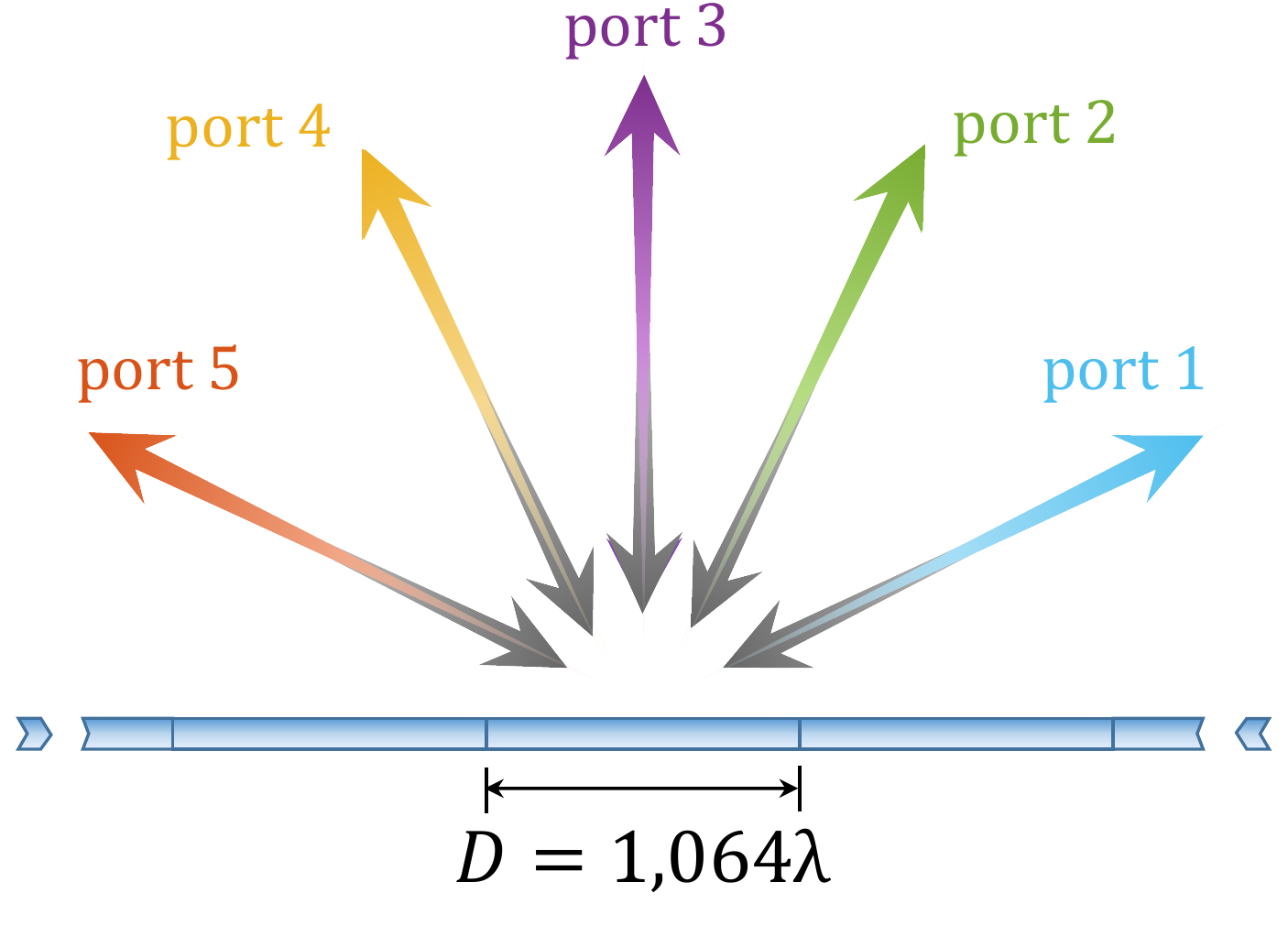}
        \caption{}
        \label{fig:1b}
    \end{subfigure}
    \caption{(a) Three-channel power splitter. A normally incident wave is reflected and split between ports 1 and 3. (b) Five-channel isolating mirror. Incident waves from different angles (the grey arrows) are reflected back to the source (the colored arrows).}\label{fig:graphs}
\end{figure}

In this work we design and verify the performance of two multi-channel devices. 
First of them is a three-channel power splitter (Fig.~\ref{fig:1a}), where a normally incident plane wave is split between ports 1 and 3, while port 2 is perfectly matched to free space (no reflections back). 
Conventional devices do not offer a possibility to control the proportion of the power for each channel, while the theory of metasurfaces allows us to design any power ratio. 
As an example, we choose the $50:0:50$ power proportion. 
To design such a surface, the general approach based on the surface impedance concept was used \cite{perfect}. The required surface impedance $Z_{\rm s}$, which is a periodic function with the period $D=\lambda/\sin\theta_{\rm r}$, reads:
\begin{equation}
Z_{\rm s}=\frac{\eta}{\sqrt{\cos\theta_{\rm r}}}
\frac{\sqrt{\cos\theta_{\rm r}}+j\sqrt{2}\sin(k_xx)}
{1-j\sqrt{2\cos\theta_{\rm r}}\sin(k_xx)},
\label{zs_splitter}
\end{equation}
where $\theta_{\rm r}$ is the reflection angle, $k_x= {2\pi \over \lambda} \sin{\theta_{\rm r}}$, and $\eta$ is the intrinsic impedance of the surrounding space.

The second multi-channel device is a five-channel isolating mirror (Fig.~\ref{fig:1b}). Retro-reflection is a well-known and widely used functionality, which can be performed by conventional devices such as blazed gratings. 
However, they are all designed for a single particular incidence configuration \cite{multi_channel}. 
In our work, for the first time the performance of retro-reflection device is extended to five channels, whereas the isolation in all of them is achieved.

\section{Experimental verification}
Fabricated samples, designed to operate at $8$ GHz, consist of arrays of rectangular metal patches of specific dimensions over a metallic ground plane (Fig.~\ref{fig:2a}). The dimensions along the $x$- and $y$-axes equal to $439$ and $262.5$~mm, respectively (for more details  see \cite{multi_channel}). The dielectric material between the patches and ground plane is Rogers 5880 ($\varepsilon_{\rm d}=2.2, \tan\delta=0.0009$) with a thickness of $1.575$ mm. The measurements were conducted in an anechoic chamber emulating the free-space environment at microwaves.

\subsection{Five-channel isolating mirror}
\begin{figure}[h!]
    \centering
        \begin{subfigure}[b]{0.20\textwidth}
        \includegraphics[width=\textwidth]{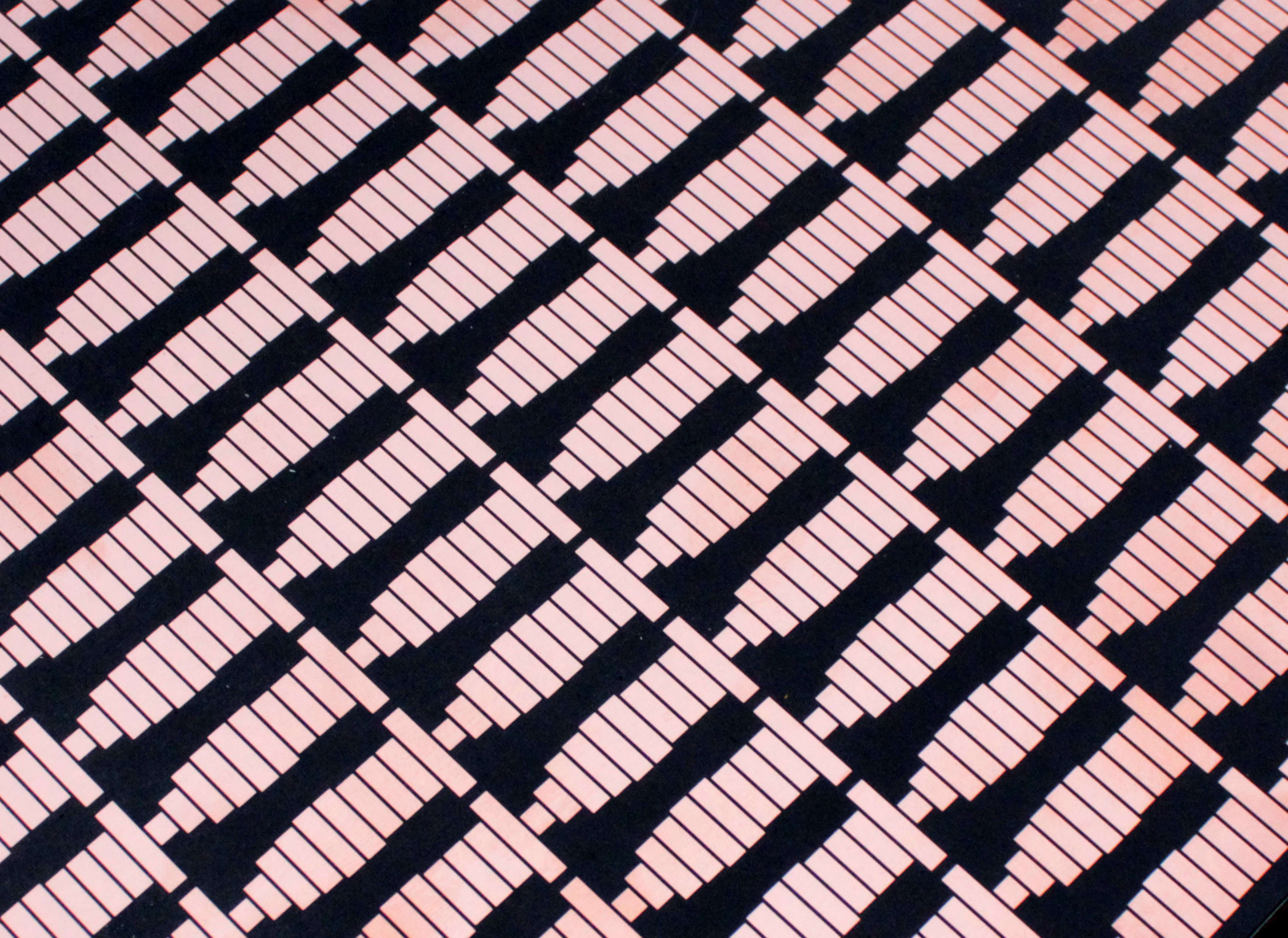}
          \caption{}
     \label{fig:2a}
    \end{subfigure}
       \hspace{0.3cm}
        \begin{subfigure}[b]{0.21\textwidth}
        \includegraphics[width=\textwidth]{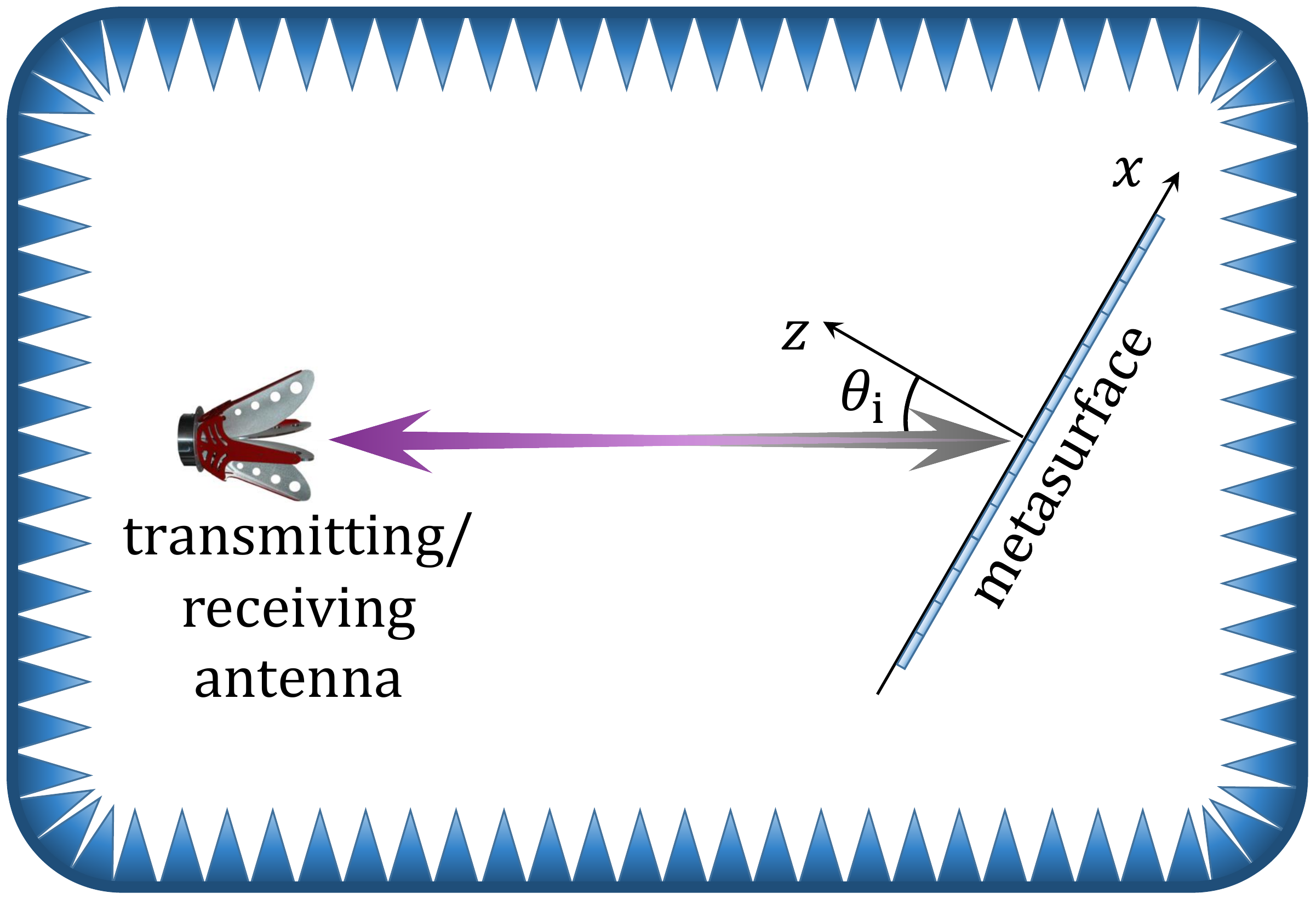}
        \caption{}
        \label{fig:3a}
    \end{subfigure}
    \hspace{0.3cm}
    \begin{subfigure}[b]{0.2\textwidth}
        \includegraphics[width=\textwidth]{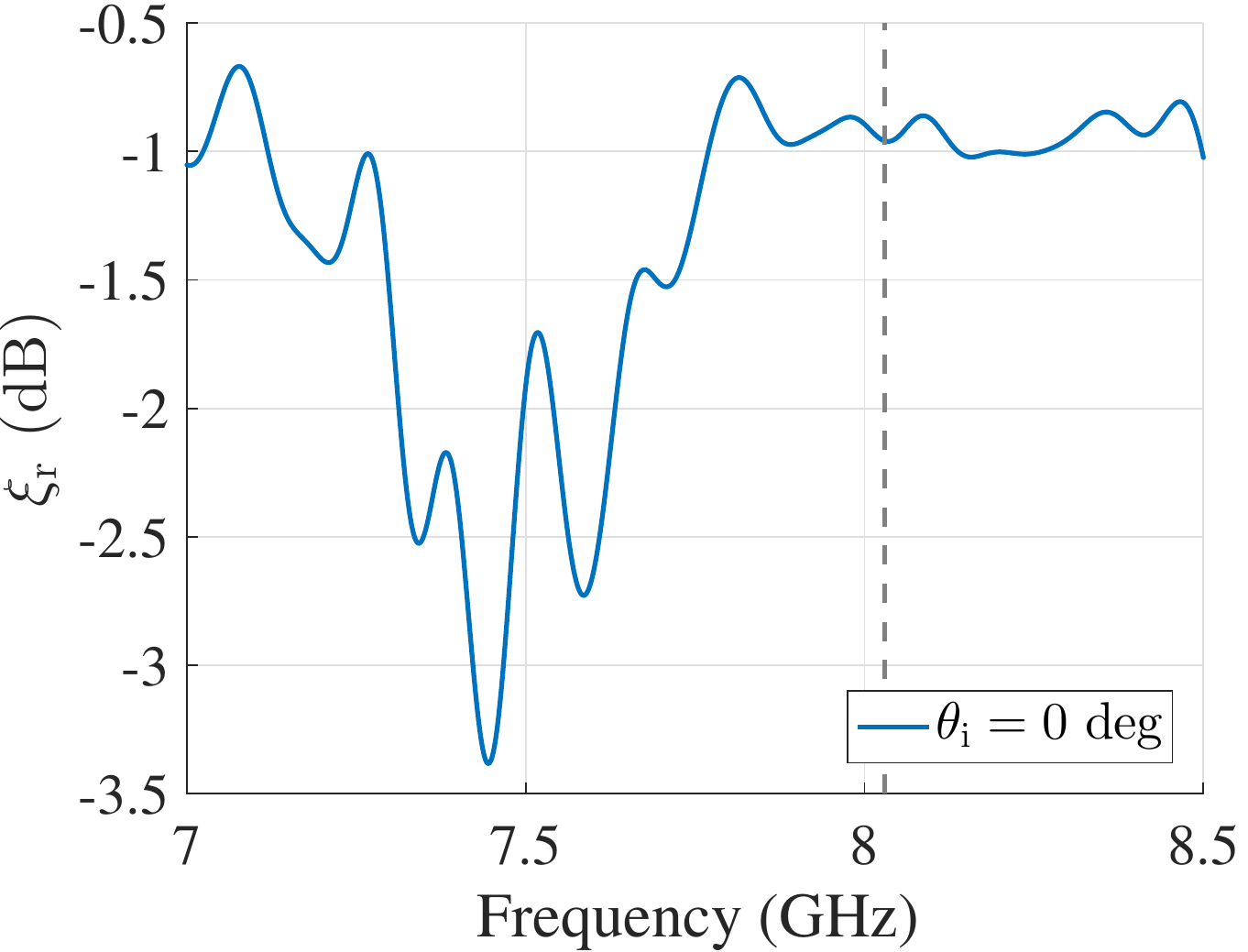}
        \caption{}
        \label{fig:3b}
    \end{subfigure}\\
    \begin{subfigure}[b]{0.2\textwidth}
        \includegraphics[width=\textwidth]{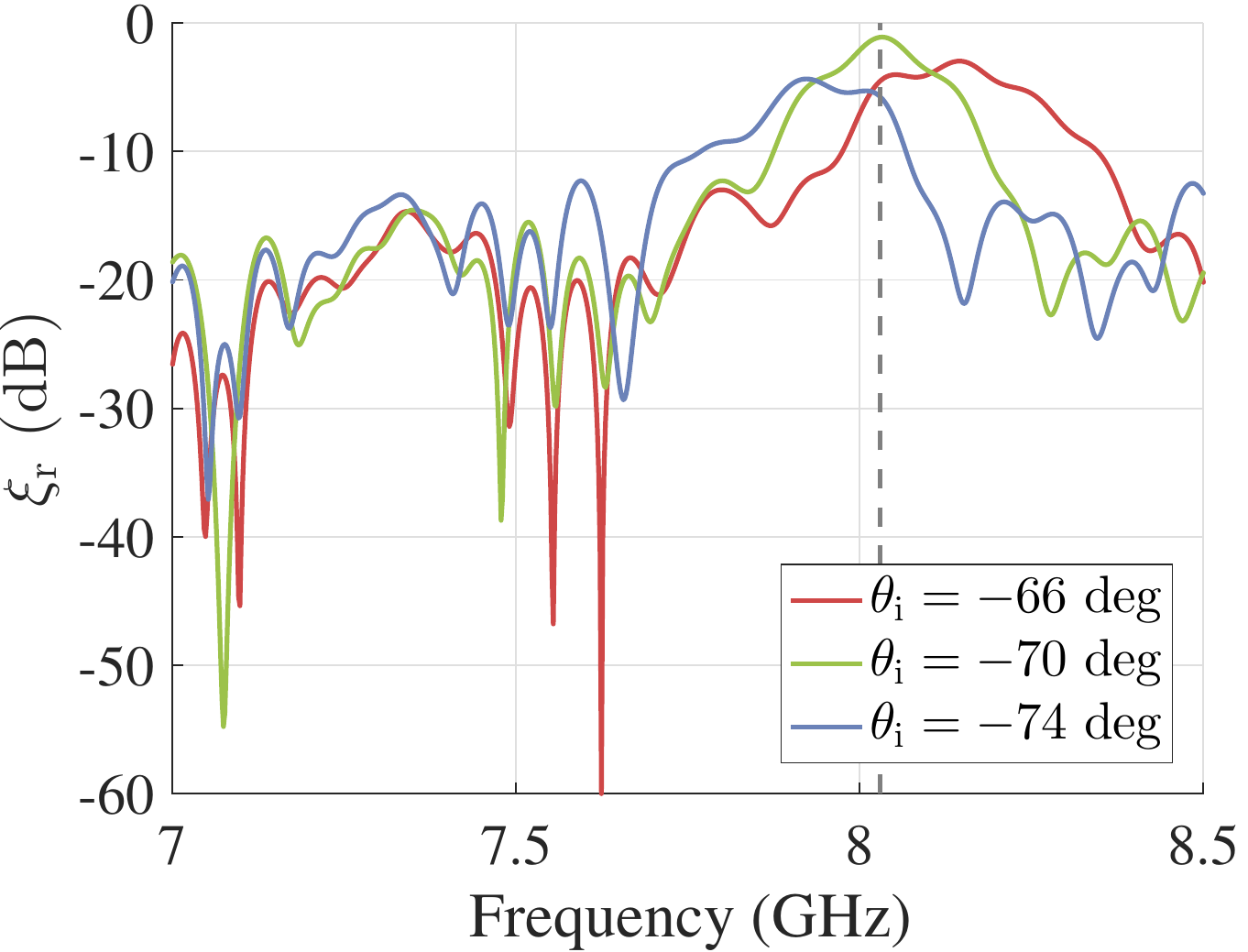}
        \caption{}
        \label{fig:3c}
    \end{subfigure} \hspace{0.3cm}
        \begin{subfigure}[b]{0.2\textwidth}
        \includegraphics[width=\textwidth]{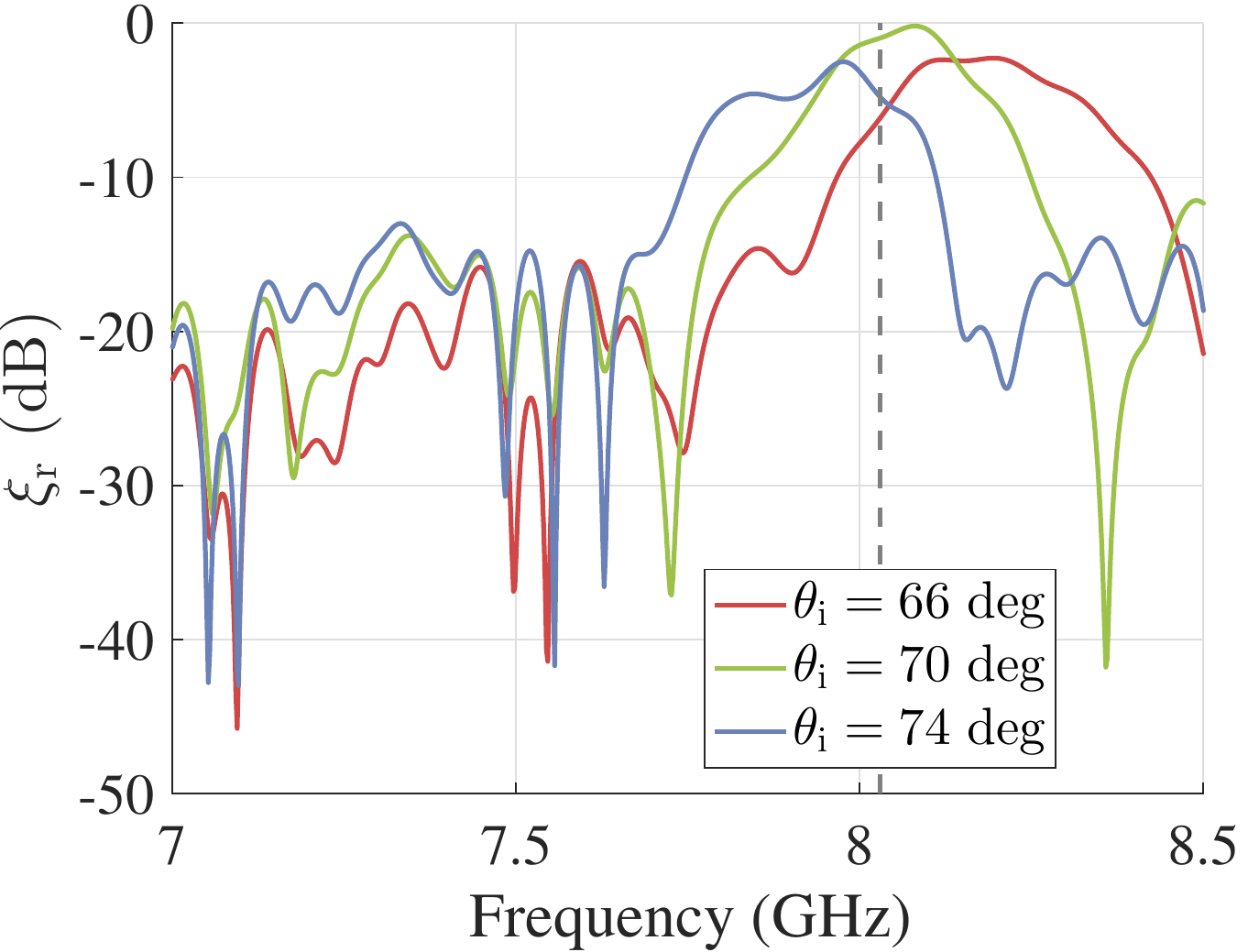}
          \caption{}
        \label{fig:3d}
    \end{subfigure}     \hspace{0.3cm}
    ~ 
    \begin{subfigure}[b]{0.2\textwidth}
        \includegraphics[width=\textwidth]{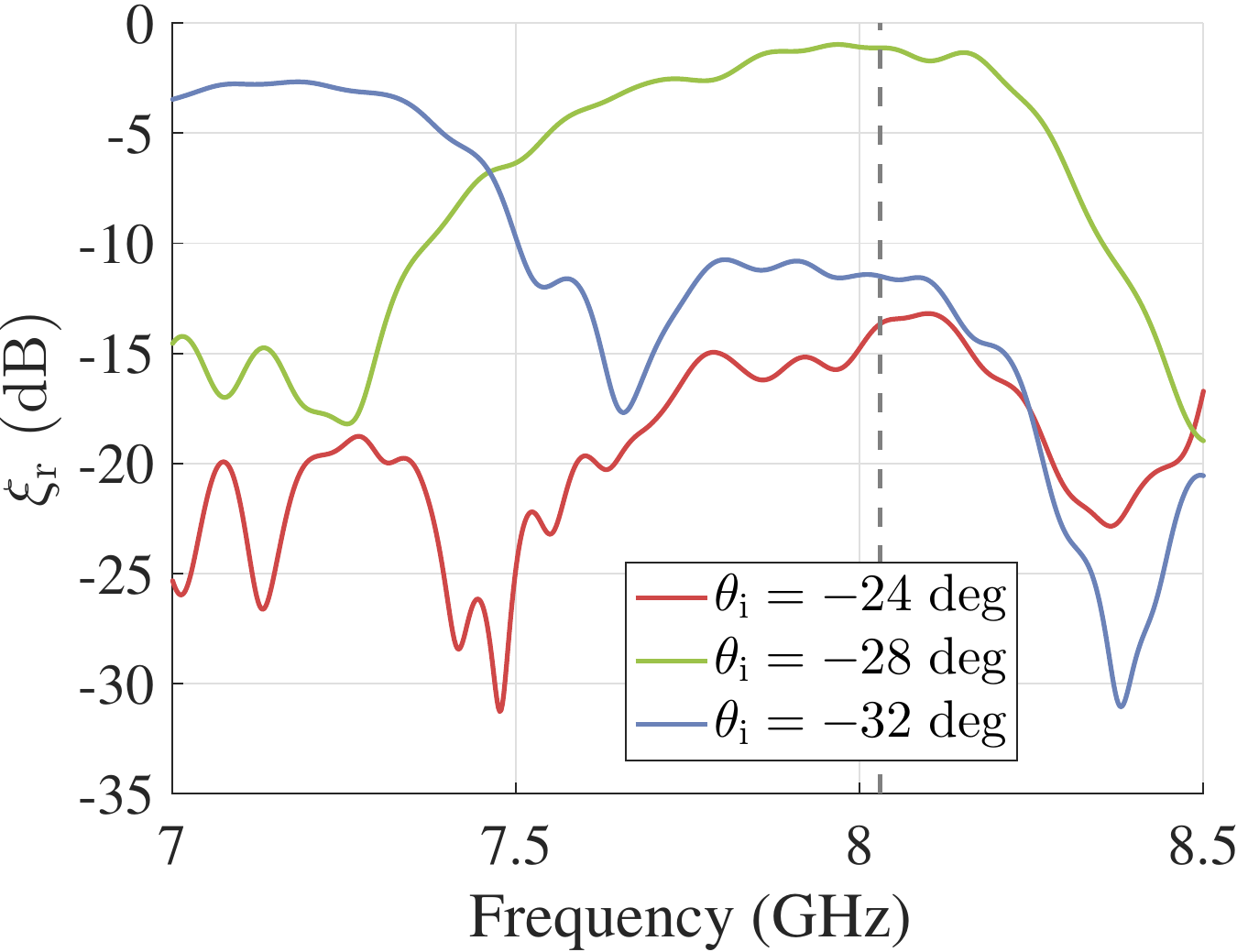}
        \caption{}
        \label{fig:3e}
    \end{subfigure}     \hspace{0.3cm}
    \begin{subfigure}[b]{0.2\textwidth}
        \includegraphics[width=\textwidth]{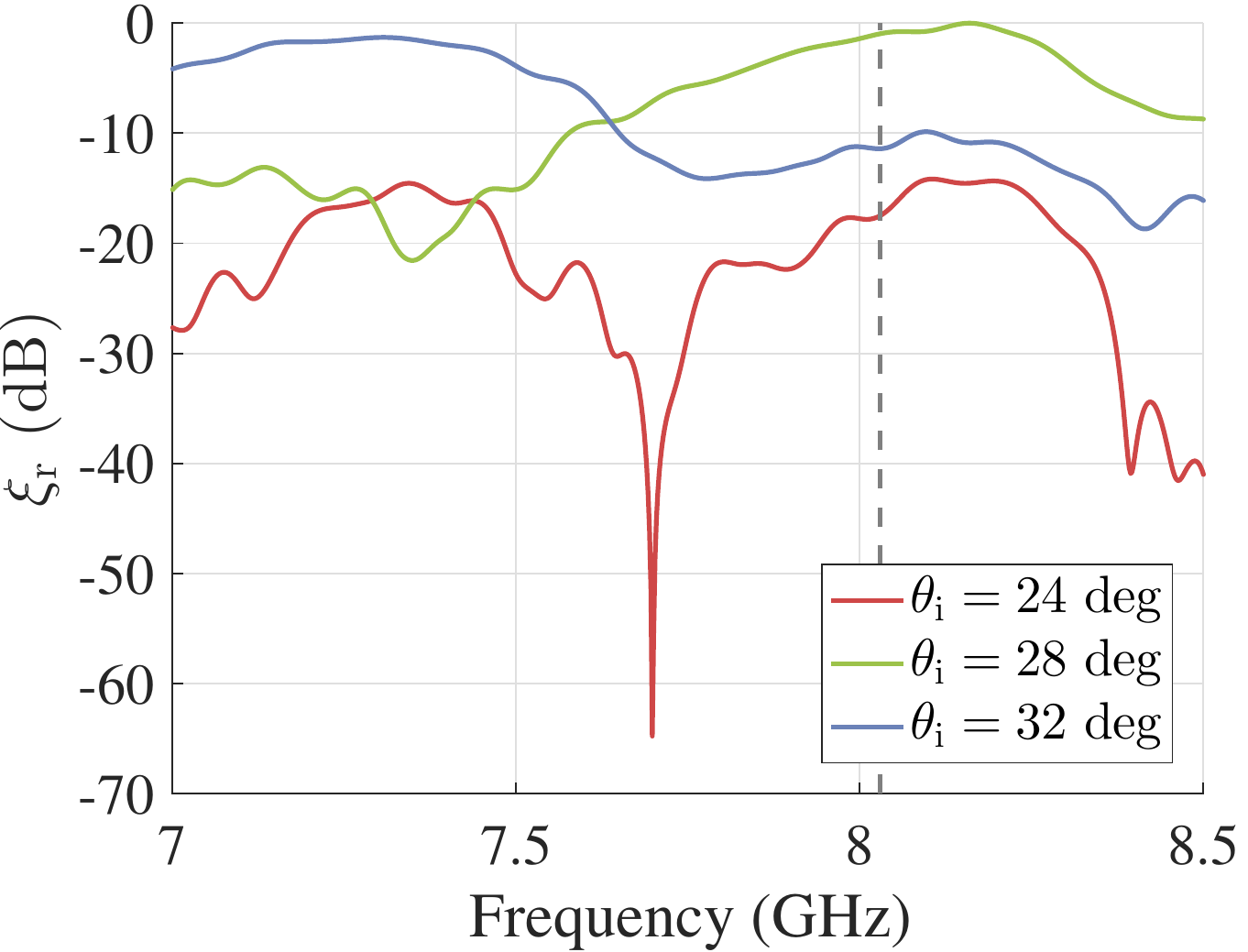}
          \caption{}
        \label{fig:3f}
    \end{subfigure}
    \caption{ (a) Fabricated metasurface. (b) Setup of the experiment, where one antenna is used for transmission and reception. Measured retro-reflection efficiency of the five-channel isolating mirror when illuminated from (c) port 3, (d) port 1, (e) port 5, (f) port 2, and (g) port 4. The operating frequency is indicated by the gray dashed line.}\label{fig:5_ports}
\end{figure}

The five-channel isolating mirror operates as a retro-reflector when the illumination angle $\theta_{\rm i}$ is $-70^{\circ}$, $-28^{\circ}$, $0^{\circ}$, $+28^{\circ}$, and $+70^{\circ}$ with an efficiency of $79.6\%$, $81.5\%$, $78.4\%$, $80.7\%$, and $87.3\%$, respectively (as numerical simulations show). 
For retro-reflection measurements the sample was placed on a rotating stand at the distance of $5.4$ m from a horn antenna used for both transmission and reception (Fig.~\ref{fig:3a}). 
The signal reflected from the plate was post-processed by a conventional time gating technique, which allows to tune the time window only for the desired signal, eliminating unwanted signals (such as parasitic reflections from the horn and the chamber walls).
The measured signals reflected from the metasurface sample are subsequently normalized by the  corresponding signal received from a reference metal plate with the same cross-section size. 
The normalized results for each channel are shown in Figs.~\ref{fig:3b}-\ref{fig:3f}.

Retro-reflection in all the channels is high and nearly equal at $8.03$ GHz. The measured efficiency at this frequency equals to $77.6\%$, $77.2\%$, $80.2\%$, $80.1\%$, and $80\%$ for illumination angles $-70^{\circ}$, $-28^{\circ}$, $0^{\circ}$, $+28^{\circ}$, and $+70^{\circ}$, respectively. Therefore, experimental and numerical results for the five-channel isolating mirror are in a good agreement.

\subsection{Three-channel power splitter}
\begin{figure}[h!]
    \centering
    ~ 
           \begin{subfigure}[b]{0.22\textwidth}
        \includegraphics[width=\textwidth]{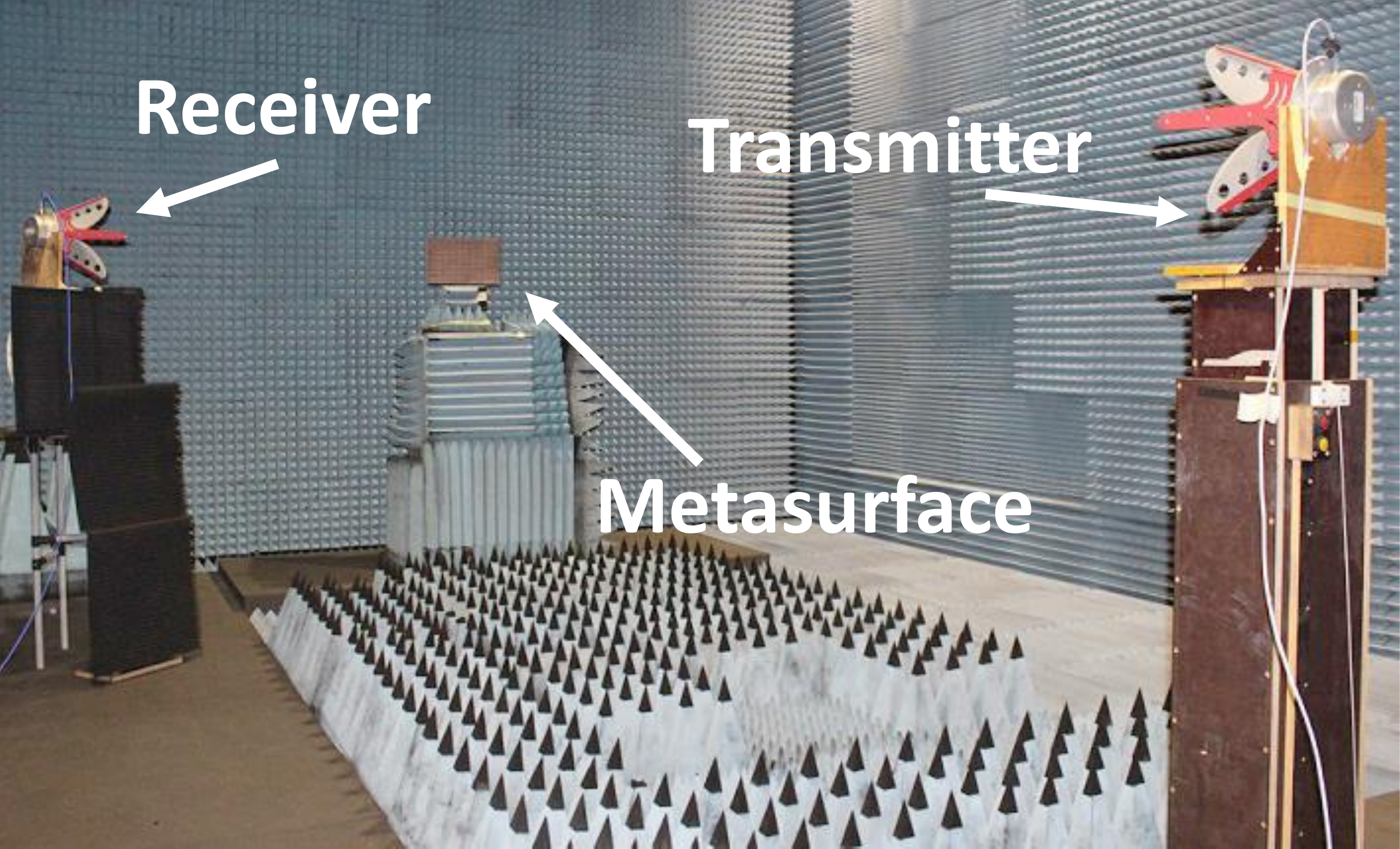}
          \caption{}
		\label{fig:2b}
    \end{subfigure}
   \hspace{0.3cm}
        \begin{subfigure}[b]{0.2\textwidth}
        \includegraphics[width=\textwidth]{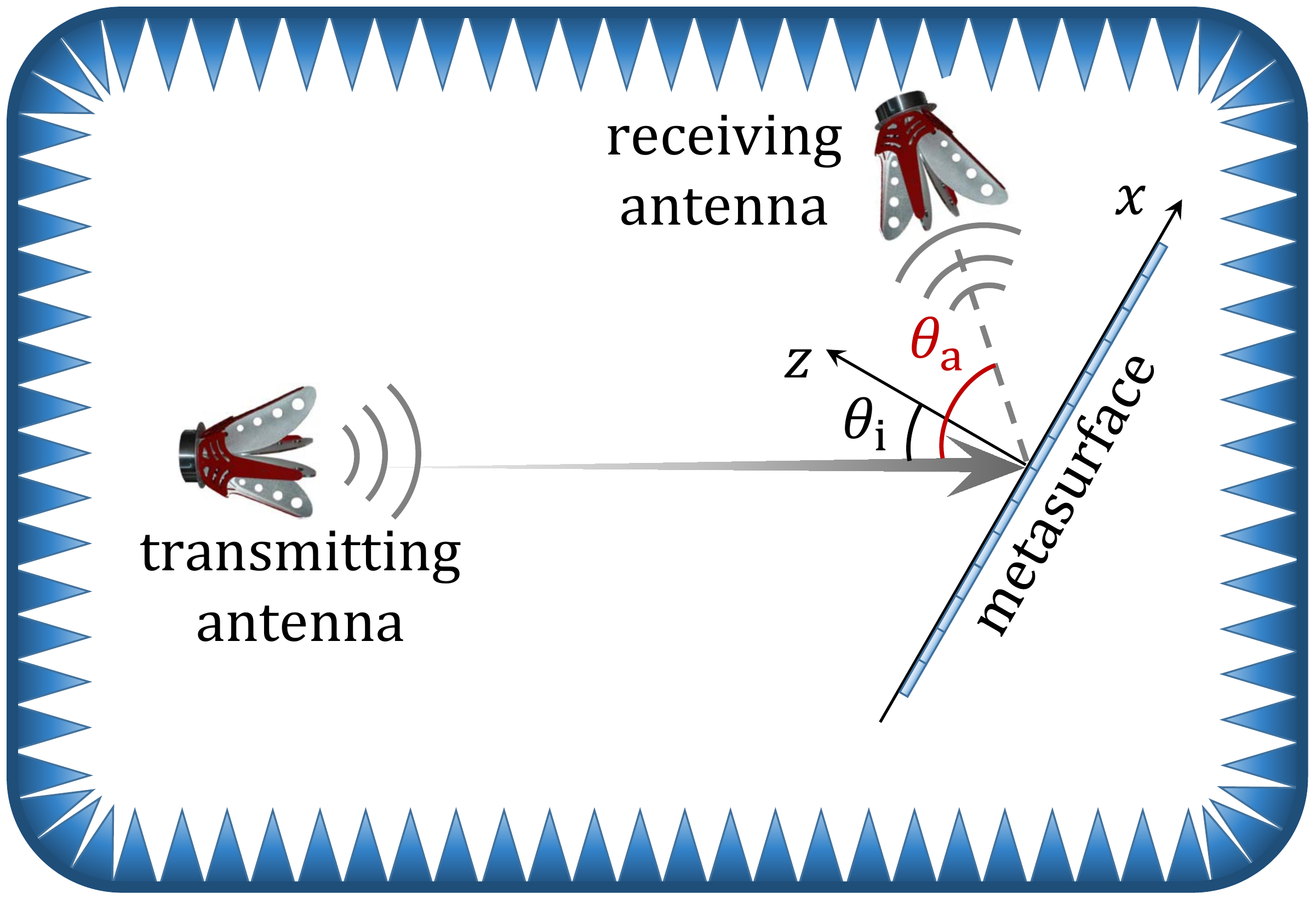}
        \caption{}
        \label{fig:2c}
    \end{subfigure}\\
        \begin{subfigure}[b]{0.2\textwidth}
        \includegraphics[width=\textwidth]{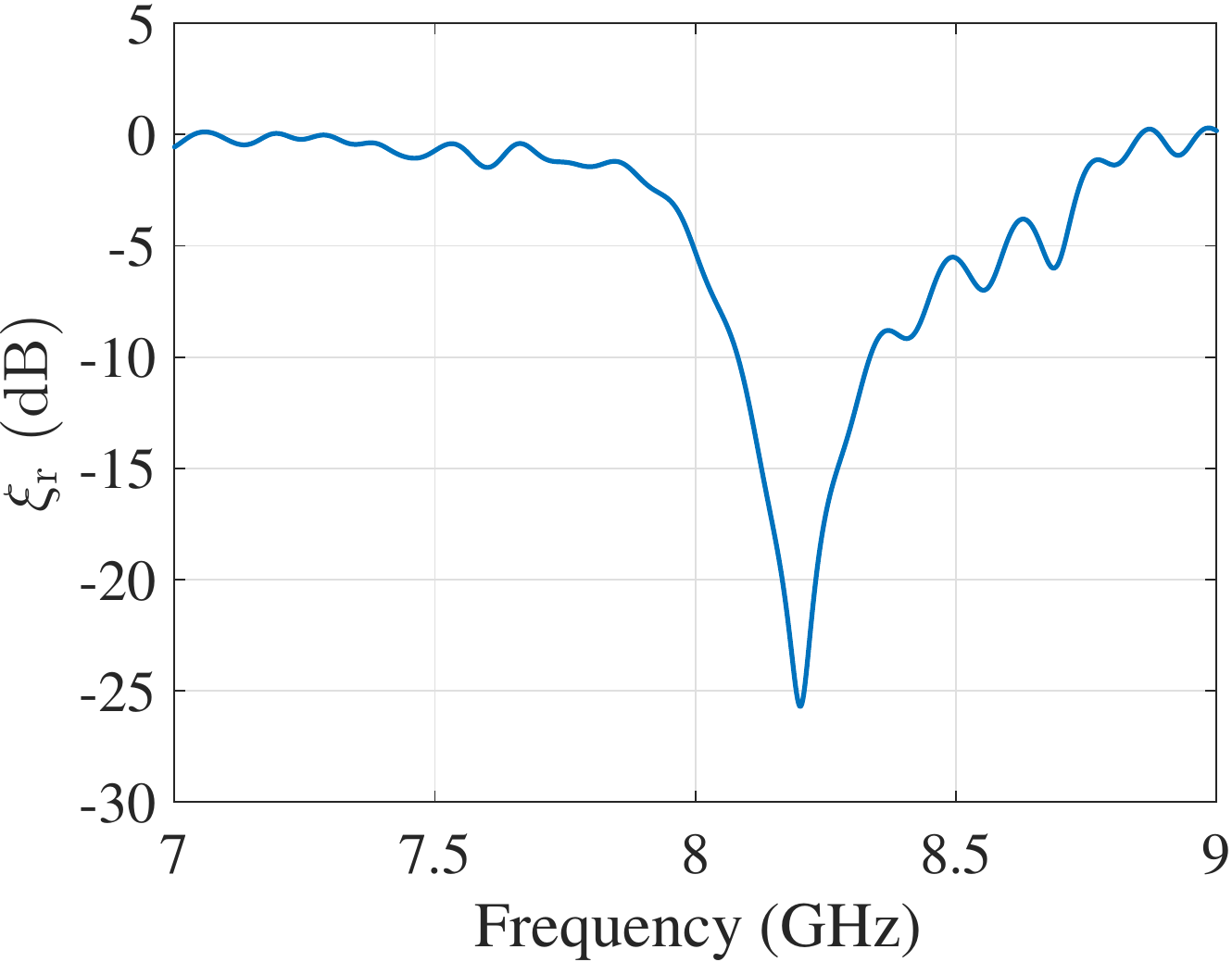}
        \caption{}
        \label{fig:2d}
    \end{subfigure} \hspace{0.3cm}
            \begin{subfigure}[b]{0.2\textwidth}
        \includegraphics[width=\textwidth]{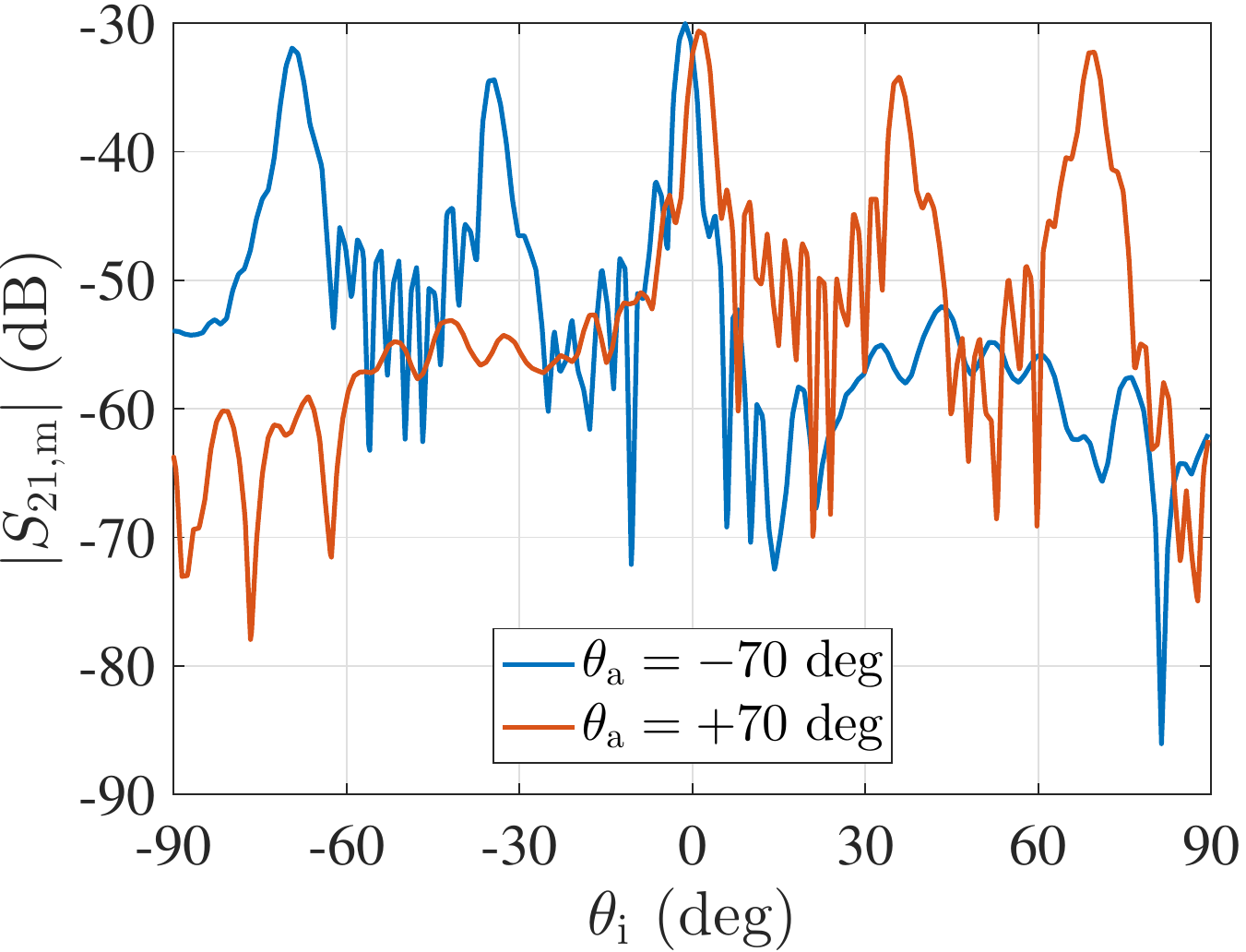}
          \caption{}
        \label{fig:2e}
    \end{subfigure}     \hspace{0.3cm}
    ~ 
    \begin{subfigure}[b]{0.2\textwidth}
        \includegraphics[width=\textwidth]{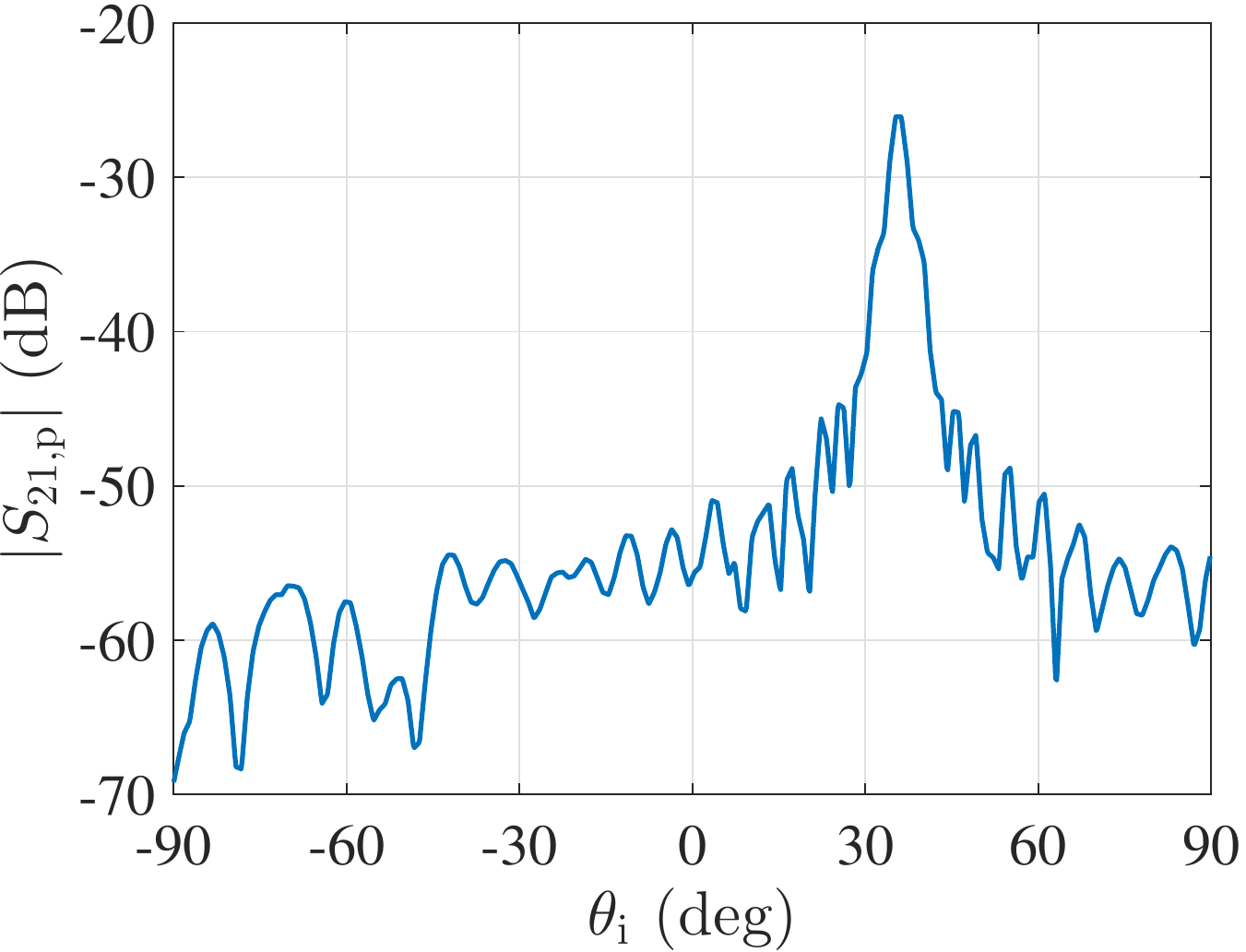}
        \caption{}
        \label{fig:2f}
    \end{subfigure}    
    \caption{Experimental setup in an anechoic chamber: (a) photo, (b) schematic top view.  (c) Reflection efficiency of the metasurface under normal incidence. Signals measured by the receiving antenna for different orientation angles $\theta_{\rm i}$ of (d) the metasurface and (e) the metal plate of the same size.}\label{fig:splitter}
\end{figure}

The designed three-channel power splitter illuminated from the normal direction splits  the energy towards $+70^{\circ}$ and $-70^{\circ}$ in proportion $49.71\%:1.06\%:45.74\%$, as numerical simulations show. 
First, retro-reflection measurements were conducted for an incidence angle $\theta_{\rm i}=0^{\circ}$ (from port 2) followed by normalization by  the signal from a metal plate of the same cross-section size (Fig.~\ref{fig:2d}). 
At a resonance frequency of $8.2$~GHz the reflected power approaches  $0.3\%$ meaning that port 2 is matched to free space, which agrees with the numerical data.
The small frequency shift can  be caused by manufacturing errors.
For power division verification, two antennas placed at distances $5.4$~m and $2.1$~m from the sample (for transmitting and receiving, respectively) were used (Figs.~\ref{fig:2b},~\ref{fig:2c}).  
The former illuminates the metasurface normally and the latter, positioned at $\theta_{\rm a}=\pm 70^{\circ}$, receives the signal reflected from the sample at the desired angle. 
The measured transmission coefficient  $|S_{21,{\rm m}}|$ for the splitter for different antenna positions is shown in Fig.~\ref{fig:2e}.
The blue curve, which has 3  peaks, corresponds to the case when the antenna is positioned at $\theta_{\rm a}=- 70^{\circ}$. The central peak ($\theta_{\rm i}=0^\circ$) is due to the reflection of the normally incident wave into port~1. 
The middle peak ($\theta_{\rm i}=-35^\circ$) corresponds to the specular reflection from the metasurface when it is  illuminated not from the main ports. The peak at $\theta_{\rm i}=-70^\circ$ is due to the excitation of the metasurface from port~1 and reflection into port~2. The red curve corresponds to the case when the antenna is positioned at $\theta_{\rm a}=+ 70^{\circ}$,  and its peaks can be explained in a similar way.

To find the power split and reflected by the metasurface towards $\theta_{\rm r}=+70^\circ$ and $\theta_{\rm r}=-70^\circ$, an additional measurement was conducted with a reference metal plate, while the  receiving antenna position was the same as used for the sample at $\theta_{\rm a}=+70^\circ$ (Fig.~\ref{fig:2f}). Using the procedure described in~\cite{ana}, the measured reflection efficiency $\xi_{\rm r}$ of the power splitter towards $\theta_{\rm r}=+70^\circ$ (for the red curve) and $\theta_{\rm r}=-70^\circ$ (for the blue curve) was calculated to be about 48\%, which is in a good agreement with the simulation results.

\section{Conclusion}
Experimental results show that it is possible to create multi-channel mirrors with required performances using properly modulated reflecting metasurfaces. Two experimentally examined samples demonstrate nearly perfect operation,  fully corresponding to numerical simulation results. 

\acknowledgement
This work was supported in part by Academy of Finland (project 287894).


\end{document}